# Tumor Connectomics: Mapping the intra-tumoral complex interaction network

Vishwa S. Parekh[a,c], Michael A. Jacobs[a,b]

[a]The Russell H. Morgan Department of Radiology and Radiological Sciences and [b]Sidney Kimmel Comprehensive Cancer Center. The Johns Hopkins University School of Medicine, Baltimore, MD 21205, USA; [c]Department of Computer Science, The Johns Hopkins University, Baltimore, MD 21208

## ABSTRACT

Tumors are extremely heterogeneous and comprise of a number of intratumor microenvironments or sub-regions. These tumor microenvironments may interact with each based on complex high-level relationships, which could provide important insight into the organizational structure of the tumor network. To that end, we developed a tumor connectomics framework (TCF) to understand and model the complex functional and morphological interactions within the tumor. Then, we demonstrate the TCF's potential in predicting treatment response in breast cancer patients being treated with neoadjuvant chemotherapy. The TCF was implemented on a breast cancer patient cohort of thirty-four patients with dynamic contrast enhanced (DCE) magnetic resonance imaging (MRI) undergoing neoadjuvant chemotherapy treatment. The intra-tumoral network connections (tumor connectome) before and after treatment were modeled using advanced graph theoretic centrality, path length and clustering metrics from the DCE-MRI. The percentage change of the graph metrics between two time-points (Baseline and 1$^{st}$ cycle) was computed to predict the patient's final response to treatment. The TCF visualized the inter-voxel network connections across multiple time-points and was able to evaluate specific changes in the tumor connectome with treatment. Degree centrality was identified as the most significant predictor of treatment response with an AUC of 0.83 for classifying responders from non-responders. In conclusion, the TCF graph metrics produced excellent biomarkers for prediction of breast cancer treatment response with improved visualization and interpretability of changes both locally and globally in the tumor.

## 1. INTRODUCTION

Complex network analysis is the study of complex, irregular and dynamic networks that are evolving in time. Complex network analysis has received considerable attention since the seminal paper by Erdos and Renyl on random graphs[1]. In the past decade, the papers by Watts and Strogatz on small world phenomenon and by Barabasi and Albert on scale free property have sparked a great amount of interest in the field of dynamically evolving complex network analysis [2, 3]. Complex network analysis techniques have shown incredible success in the analysis of social networks, world wide web, brain networks, and others [4-8].

In this work, we apply and extend the application of complex network analysis to intra-tumoral network analysis based on radiological imaging of cancer. Tumors are extremely heterogeneous and comprise of a large number of tumor microenvironments or clusters. Different tumor microenvironments interact with each based on complex high-level relationships which could provide important insight into the organizational structure of the tumor network. In addition, evaluating how the complex network structure evolves with time or treatment could provide metrics for evaluation of treatment response.

The current state-of-the-art methods for extraction of intra-tumoral heterogeneity from radiological images are based on statistical texture analysis, which provide information about the intensity distribution and inter-pixel relationships within regions of interest drawn on radiological images [9, 10]. These methods, however, do not attempt to uncover the complete underlying network structure of the tumor and the relationship with its surroundings. In addition, visualization and interpretation of traditional texture analysis methods can be difficult. On the other hand, representing the tumor as a network structure provides increased interpretability and visualization of the tumor substructure. In this manuscript, we develop the tumor connectomics framework for complex network analysis of breast tumors using multiparametric (mp) MRI and demonstrate its performance for prediction of treatment response to neoadjuvant chemotherapy breast cancer.

## 2. MATERIALS AND METHODS

### 2.1 Clinical data

We developed the tumor connectomics framework (TCF) for treatment response prediction of breast cancer on an independent patient dataset obtained from the University of California San Francisco (UCSF) ISPY study [11, 12]. The ISPY study consisted of thirty-four patients with mp MRI who underwent neoadjuvant chemotherapy. Treatment response was defined by clinical standards and used for ground truth. The clinical response metric (CR) is defined in four categories – 1. no evidence of disease (CR=1), 2. >1/3 decrease in clinically longest diameter (LD) (CR=2), 3. <1/3 decrease in LD (CR=3) and progressive disease (PD) (CR=4). The tumor connectomics framework was applied at the baseline and after the first cycle (D7) of the drug and correlated to clinical response metrics (C1-C4).

### 2.2 MRI protocol

The MRI scans were obtained on a 1.5 T magnet in the sagittal using a dedicated breast coil. The MR images obtained were fat suppressed, T1 weighted dynamic contrast enhanced series obtained unilaterally in the sagittal orientation with TR≤20ms, TE = 4.5ms, FA≤45°, FOV: 16-18 cm, matrix size > 256x192, ST ≤ 2.5 mm.

### 2.3 Tumor Connectomics Framework

**Figure 1** illustrates a typical tumor connectome modeled from a DCE-MRI dataset with the subtracted DCE-MRI shown for visualization of the lesion and the connections. The goal of the TCF is to transform the mp radiological imaging dataset consisting of D images into a TCF map where each voxel position is overlaid with different graph metrics computed for that voxel position within the lesion (**Figure 2**). Mathematically, the mp radiological imaging dataset can be represented as $X = \{x_{11}, x_{12}, ..., x_{mn}\} \subset R^D$ where, $x_{ij}$ is the tissue signature at voxel position (i,j), m and n are the number of rows and columns in each image in the mp dataset and D is the number of images in the high dimensional space. The mp radiological imaging dataset, X, is transformed into a TCF map, $Y = \{y_{11}, y_{12}, ..., y_{mn}\} \subset R$ using the following procedure:

1. In the first step, pairwise Euclidean distance matrix is computed for the *D* dimensional image space.
2. The second step involves identifying the nearest neighborhood of every point in the high dimensional space, X.
3. The third step involves transforming the Euclidean distance matrix into a geodesic distance matrix and an adjacency matrix using the nearest neighborhood information extracted from step 2.
4. The fourth step involves overlaying the connections obtained from step 3 on the anatomical image for visualization of tumor microenvironments and functional connections.
5. In the next step, the centrality and clustering graph metrics are extracted from the geodesic distance matrices computed in step 3.
6. The final step involves overlaying each graph metric visualization on the anatomical image to create a TCF map.

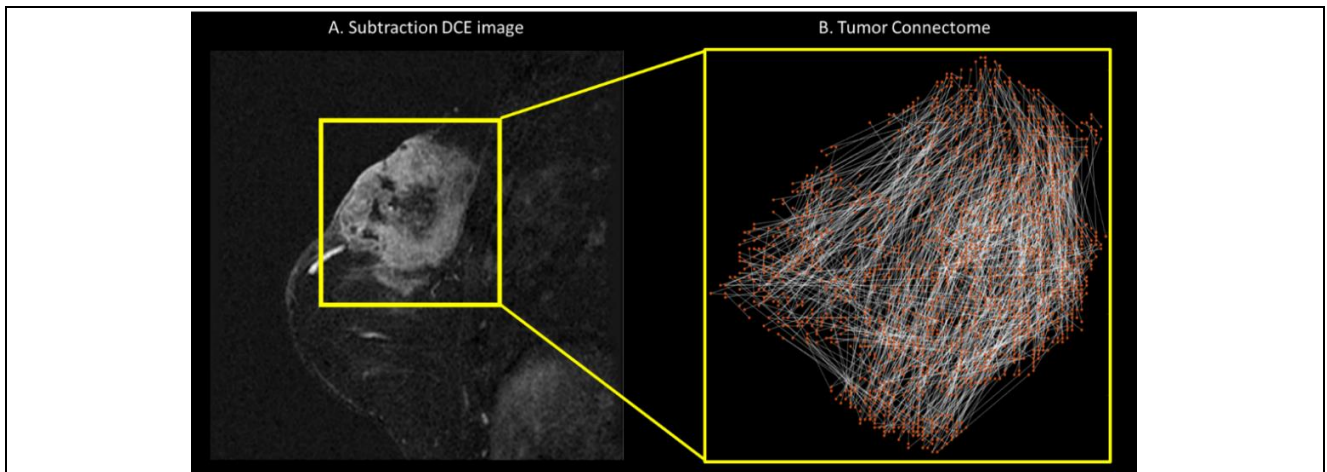

**Figure 1**. Illustration of the tumor connectome modeled for an example DCE-MRI dataset with subtraction DCE-MRI shown for visualization. (a) Subtraction dynamic contrast enhanced (DCE) MRI image (b) Intra-tumoral network or the tumor connectome

For calculating the geodesic distances (GD), all the images were normalized to the range zero to one. The GD was further normalized to the range zero to one for mp radiological imaging datasets by dividing GD with $\sqrt{D}$, where D = dimensionality of the dataset. The threshold for computing the GD was chosen based on grid search optimization. All the graph metrics were defined based on the traditional definition [5]. One exception was the average path length (APL), where the APL for any node (i,j) was defined as the mean of the shortest path distances originating from the node (i,j) to every other node reachable from the node (i,j). The rationale behind the change in the APL definition was that if a node is not reachable to the rest of the network, the original APL definition would result in most nodes having an APL of infinity.

## 2.4 Statistical analysis

Summary statistics (mean and standard error of the mean) were recorded for the graph metrics. The percent differences in the graph metrics were used to predict treatment response using the tumor connectome. Student's t-test was performed to compare the different classes of clinical response. We computed the area under the receiver operating characteristic (ROC) curve (AUC) for each of the graph metrics. Statistical significance was set at $p<0.05$.

## 3. RESULTS

The tumor connectomics framework (TCF) was tested on a breast cancer patient cohort of thirty-four patients. Of the thirty-four, eight patients had clinical response, CR=1, seventeen patients had CR=2, seven patients had CR=3 and two patients had CR=4. **Figure 2** illustrates the TCF maps for degree centrality, betweenness centrality and clustering coefficient for an example responder (CR=1) and an example non-responder (CR=3).

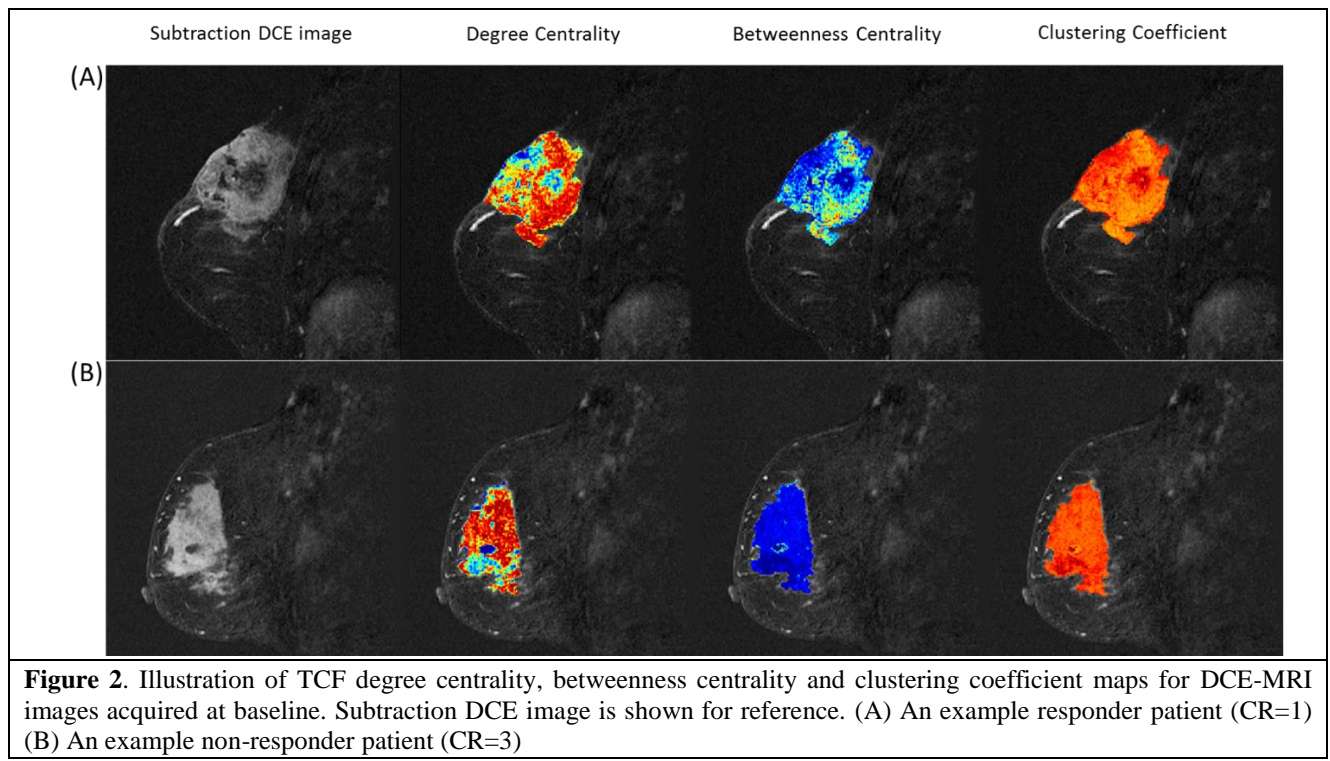

**Figure 2**. Illustration of TCF degree centrality, betweenness centrality and clustering coefficient maps for DCE-MRI images acquired at baseline. Subtraction DCE image is shown for reference. (A) An example responder patient (CR=1) (B) An example non-responder patient (CR=3)

The most predictive graph metric based on the Student's t-test between responders (CR=1,2) and non-responders (CR=3,4) was degree centrality. The optimal value for the threshold for geodesic distance computation was computed by grid search optimization using the p-value from t-test on degree centrality as the evaluation metric. The optimal value for threshold was found to be 0.1 as illustrated in **Figure 3**. The corresponding TCF degree centrality maps for a responder (CR=1) and a non-responder (CR=3) patient have been illustrated in **Figure 4**. The TCF maps are color coded with synthetic colors from blue to red, with blue corresponding to low degree centrality and red corresponding to high degree centrality. The degree centrality for the responder decreased with treatment compared to the non-responder and is evident from the increase in low degree centrality areas (blue) for the responder (**Table 1**).

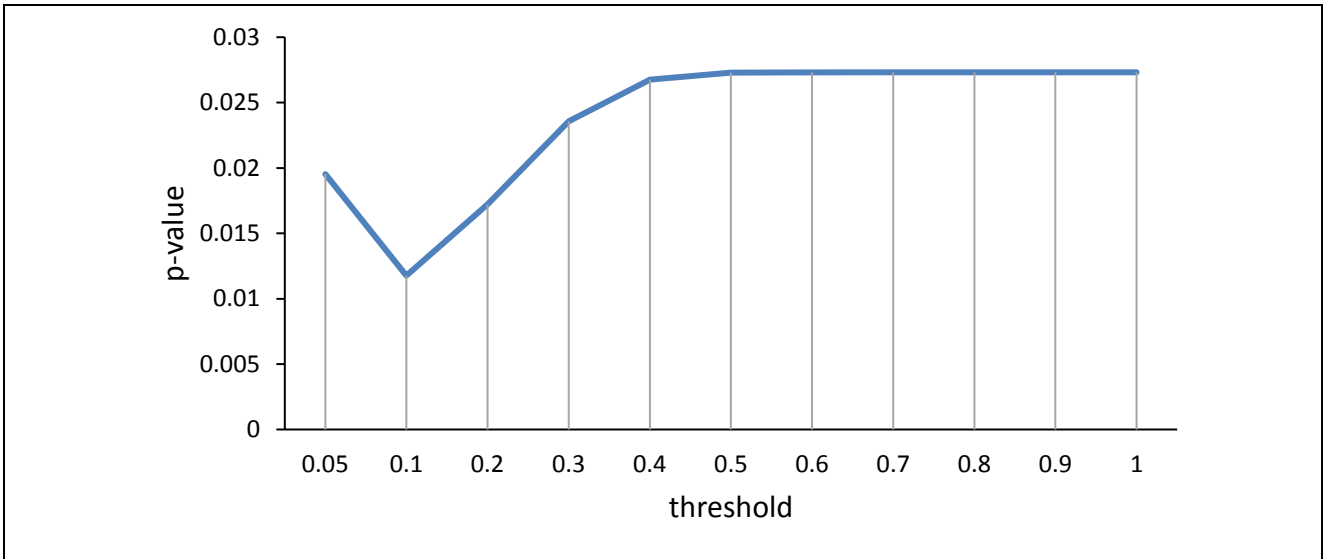

**Figure 3.** Plot of delta degree centrality p-value vs. geodesic distance threshold. It can be seen from the curve that the minimum occurs at the threshold value of 0.1

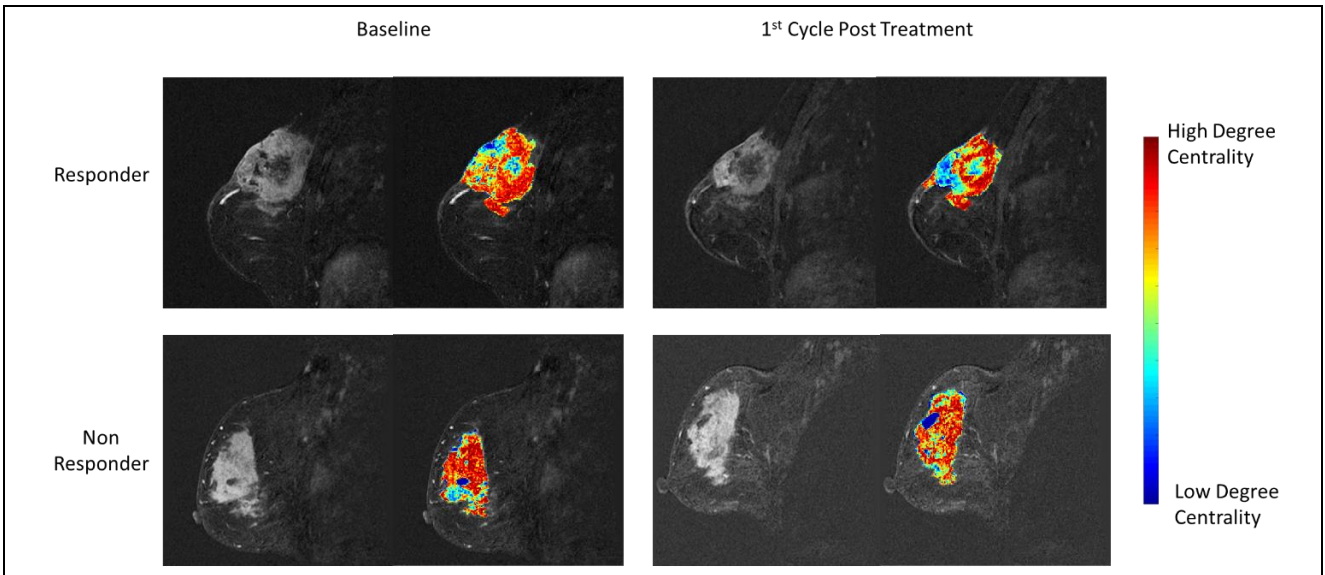

**Figure 4.** Temporal evolution of the degree centrality metric for an example responder and a non-responder. The color coding is displayed on the right. The figure demonstrates the increase in low degree centrality regions (blue) for the responder as compared to the non-responder.

The degree centrality and delta degree centrality (percentage difference between the degree centrality values) for the four response groups have been tabulated in **Table 1**. **Figure 5(a)** illustrates the evolution of degree centrality with treatment for all the four patient groups as a bar graphs. The degree centrality for the responders, CR=1 and CR=2 significantly ($p<0.05$) decreased after treatment, where the degree centrality for the non-responders did not change after treatment. The bar graph illustrating the changes in the Response Assessment Criteria (longest lesion diameter) after treatment is shown in **Figure 5(b)** for comparison. Finally, the AUC for distinguishing between the responders (CR=1,2) and non-responders (CR=3,4) was obtained at 0.83. The ROC curve is shown in **Figure 6**. The sensitivity and specificity for predicting treatment response were obtained at 80% and 78%.

**Table 1**. The tumor connectomics network degree centrality values obtained from pre-treatment MRI and MRI acquired after first cycle of treatment and the percentage difference between the two-degree centrality measurements (delta degree centrality).

| Clinical Response | Degree Centrality (Pre-Treatment) | Degree Centrality (1st Cycle Post Treatment) | Delta Degree Centrality |
|---|---|---|---|
| CR=1 | 1171.74±454.96 | 471.82±169.82 | -0.74±0.20 |
| CR=2 | 1237.04±164.17 | 684.29±115.52 | -0.47±0.14 |
| CR=3 | 782.94±584.52 | 767.75±162.47 | 0.03±0.14 |
| CR=4 | 1507.87±622.11 | 1471.80±258.42 | -0.04±0.14 |
| P value (CR=1,2 vs CR=3,4) | 0.49 | 0.28 | 0.01 |

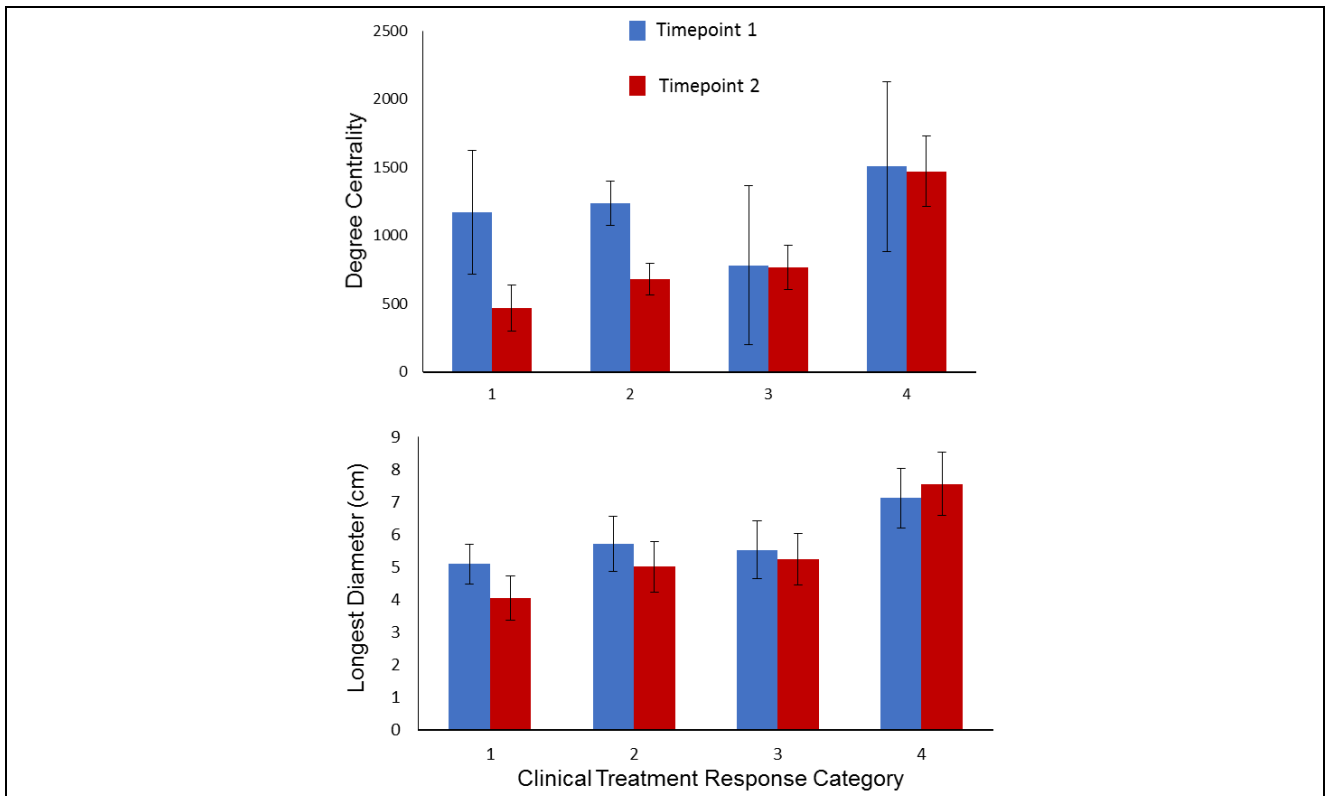

**Figure 5.** A bar plot demonstrating the delta change in the average degree centrality of the tumor connectome calculated from the DCE-MRI and the longest lesion diameter acquired pre and post (1st cycle) treatment. The degree centrality decreased for the patients with CR=1 and CR=2 and stayed the same for the patients with CR=3 and CR=4. On the other hand, the longest diameter decreased for the patients with CR=1, CR=2 and CR=3 and increased for the patients with CR=4.

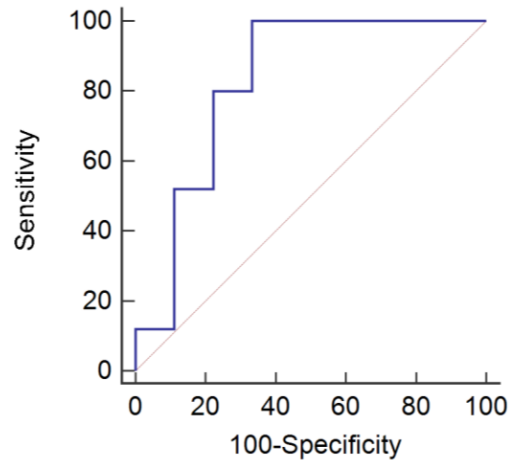

**Figure 6.** The receiver operating characteristic (ROC) curve for prediction of treatment response in breast cancer patients receiving neoadjuvant chemotherapy using degree centrality graph metric. The area under the ROC curve (AUC) was obtained at 0.83 with sensitivity and specificity values of 80% and 78% respectively.

## 4. DISCUSSION

We have developed a tumor connectomics framework (TCF) for visualization and analysis of intra-tumoral complex network using radiological imaging. The TCF demonstrated excellent values for sensitivity, specificity and AUC for prediction of treatment response in breast cancer patients being treated with neoadjuvant chemotherapy. The tumor connectomes at the baseline and after the 1$^{st}$ cycle of treatment had significantly different structural properties. The patient's response to treatment was characterized by destruction of the inter-voxel connections as represented by the network degree centrality. These changes in the tumor microenvironment could be used as potential response biomarkers during chemotherapy. A tumor consists of multiple sub-regions or tumor microenvironments that interact with each other in a complex network model. The TCF produced a visualization of the temporal evolution of the intra-tumoral network connections which could allow us to identify and map specific changes within each tumor subregion or microenvironments with treatment. There are certain limitations to this study. This was a preliminary study aimed at evaluating the feasibility and efficacy of the newly developed tumor connectomics framework in predicting treatment response in breast cancer patients being treated with neoadjuvant chemotherapy. In the future, this framework would need to be implemented and validated in a larger patient cohort and also evaluated for different diagnostic and prognostic applications.

In conclusion, the TCF metrics produced novel biomarkers for prediction of treatment response in breast cancer with improved visualization and interpretability of changes both locally and globally in the tumor.

## 5. ACKNOWLEDGEMENTS

National Institutes of Health grant numbers: 5P30CA006973 (IRAT), U01CA140204, and 1R01CA190299 and The Tesla K40 used for this research was donated by the NVIDIA Corporation.